\documentclass[aps,pre,twocolumn,groupedaddress]{revtex4-1}
\usepackage{graphicx}
\usepackage{color}
\usepackage{amsmath}
\usepackage{amssymb}
\usepackage{mathrsfs}
\usepackage{subcaption}

\newcommand\dd{\mathrm{d}}

\DeclareMathOperator{\sech}{sech}

\begin{document}

\title{Koopman analysis of isolated fronts and solitons}

\author{Jeremy P. Parker}\email{jeremy.parker@damtp.cam.ac.uk}
\author{Jacob Page}\email{jacob.page@damtp.cam.ac.uk}

\affiliation{Department of Applied Mathematics and Theoretical Physics,
Centre for Mathematical Sciences, University of Cambridge, Wilberforce Road, Cambridge CB3 0WA, UK}

\date{\today}

\begin{abstract}
    A Koopman decomposition of a complex system leads to a representation in which nonlinear dynamics appear to be linear.
    The existence of a linear framework with which to analyse nonlinear dynamical systems
    brings new strategies for prediction and control,
    while the approach is straightforward to apply to large datasets owing to a connection with dynamic mode decomposition (DMD).
    However, it can be challenging to connect the output of DMD to a Koopman analysis since there are relatively few
    analytical results available, while the DMD algorithm itself is known to struggle in situations involving the propagation of a localised
    structure through the domain. 
    Motivated by these issues, we derive a series of Koopman decompositions for localised, finite-amplitude solutions of classical nonlinear PDEs.
    We first demonstrate that nonlinear travelling wave solutions to both the Burgers and KdV equations have \emph{two} Koopman decompositions;
    one of which converges upstream and another which converges the other downstream of the soliton or front.
    The crossover point in space-time where both expansions break down is typically at the centre of the isolated structure itself.
    We then use the inverse scattering transform to derive a full Koopman decomposition for (pure soliton) solutions to the KdV equation, identifying
    Koopman eigenvalues, eigenfunctions and modes.
    Our analysis indicates that there are many possible Koopman decompositions when the solution involves the interaction of multiple solitons.
    The correct expansion to use at any point in space (i.e. the one that converges) depends on the relative positions of all solitons,
    and remains valid until a soliton passes through the observation point.
    The existence of multiple expansions in space and time has a critical impact on the ability of DMD to extract Koopman eigenvalues and modes -- which must be performed within a
    temporally and spatially localised window to correctly identify the separate expansions.
    In addition, we provide evidence that these features may be generic for isolated nonlinear structures by applying DMD to a moving breather solution of the sine-Gordon equation.
\end{abstract}


\maketitle

\section{Introduction}
The recent emergence and development of Koopman analysis for studying dynamical systems, starting with
\citet{Mezic2005}, is a major step towards the often sought-after goal of being able to understand
complex behaviour in nonlinear dynamical systems as a combination of much simpler behaviours. The Koopman
operator \citep{Koopman1931} is a \emph{linear} operator acting on the space of observables for potentially \emph{nonlinear}
systems, allowing us to perform spectral decompositions in the usual way \citep{Rowley2009,Mezic2013}. 
The resulting Koopman decompositions (or expansions) of observables, and in particular the state of the system, cast the evolution as a
sum of spatial Koopman modes with exponential temporal behaviour.
This is possible via a projection of the observable of interest onto Koopman \emph{eigenfunctions}, scalar functionals of the state of the system which have a `linear' evolution despite the underlying nonlinear dynamics.
In this perspective, the fixed Koopman modes assume a secondary importance despite their physical significance, and can be regarded as the coefficients in the expansion \citep{Rowley2009,Mezic2013}.

One particularly attractive feature of Koopman analysis is the apparent simplicity with which it can be applied in large-scale problems, such as fluid turbulence, owing to a connection with dynamic mode decomposition (DMD) \citep{Schmid2010}.
DMD is a numerical technique that identifies a least-squares best-fit linear operator that maps between equispaced-in-time snapshots,
and which can be used to represent a time series as a sum of spatial modes with exponential dependence on time.
In a series of important contributions, various authors have identified strict requirements under which DMD is capable of performing a Koopman mode decomposition \citep{Rowley2009,Tu2014,Williams2015,Brunton2016,Rowley2017}.


The DMD algorithm is straightforward to apply to very complex systems since it requires only a sequence of snapshot pairs as input.
In particular, it has seen many applications
in fluid dynamics \citep{Schmid2010a,Jovanovic2014,DMDkutz},
though it is also increasingly being used in other areas, for example in neuroscience \citep{BruntonB2016}. 
However, it is often difficult to verify that the low-rank dynamics identified in DMD correspond to a Koopman decomposition
due to a lack of analytical results beyond ODE model problems \citep[e.g.][]{Bagheri2013,Brunton2016,Rowley2017}.
Some of these ODE results have allowed extraction of Koopman modes in more complex nonlinear PDEs, e.g. the Stuart-Landau equation examined by \citet{Bagheri2013} describes the transient collapse of unstable flow past a cylinder onto a limit cycle,
and this connection allowed him to find the corresponding Koopman modes for the velocity field.
Certain nonlinear PDEs can also be rendered linear under a transformation of the state variable which allows for identification of Koopman eigenvalues \citep[e.g.][]{Page2018,Kutz2018}.
\citet{Page2018} exploited the linearising transform to derive a full Koopman decomposition for the velocity field in the Burgers equation.
In this work we exploit a similar feature in the KdV equation to derive Koopman decompositions there.

Beyond DMD, a variety of alternative methods to extract Koopman decompositions have been proposed.
For example, \citet{Sharma2016} have found a connection between Koopman modes and the `response modes' of the resolvent operator.
In statistically stationary flows, \citet{Arbabi2017} have demonstrated an approach motivated by signal processing to allow for extraction of Koopman modes and eigenfunctions.
Other approaches involve altering the snapshots on which DMD is applied, by adding additional functionals (observables) of the state of the system \citep{Williams2015} or by `stacking' snapshots of the state equispaced-in-time along the trajectory to form a single large observable \citep{Brunton2017}.

However, despite this progress there are still open questions as to how Koopman and DMD should be applied to systems which transit between multiple simple invariant solutions \citep{Brunton2016,Page2019}.
In fact, Koopman analysis applied to a simple ODE with a pair of fixed points \citep{Page2019} has shown that each simple invariant solution has an associated Koopman expansion.
Each expansion is convergent up to a crossover point along the heteroclinic connection between the fixed points.
This introduces a critical constraint on DMD, which to function as a proxy for Koopman must be performed on an observation window in which there is a single valid decomposition.
In addition, it is known that the DMD algorithm struggles when applied to localised travelling waves \citep[e.g.][]{DMDkutz} both in providing a low rank approximation to the dynamics and in extrapolating beyond the observation window.
Our analysis of the KdV equation suggests that these two behaviours may be related, as we show that localised nonlinear waves possess multiple Koopman decompositions, each of which converges in different regions of space-time.
For DMD to extract the different expansions, observations must be restricted in both time and space to a region where a single expansion holds.

The remainder of this paper is structured as follows.
In \S \ref{sec:koopmanintro} we introduce the Koopman operator and derive a pair of Koopman decompositions for a travelling-front solution of the Burgers equation.
In \S \ref{sec:kdv} we perform a similar analysis for a one-soliton solution of the KdV equation,
before using the inverse scattering transform to derive Koopman eigenfunctions, eigenvalues and modes
for general (non-dispersive) solutions to the KdV equation, establishing the need for potentially many different Koopman decompositions in a generic case.
The consequences of these decompositions for DMD are examined in \S \ref{sec:dmd}, and an observable that can robustly determine Koopman eigenvalues and modes is defined.
We then apply DMD to find Koopman decompositions of the sine-Gordon equation, where the analytical decomposition is unknown.
Finally, concluding remarks are provided in \S \ref{sec:conclusion}.

\section{Koopman decompositions of nonlinear dynamics}
\label{sec:koopmanintro}
In this paper we will consider nonlinear PDEs of the form
\begin{equation}
    \partial_t u(x,t) = F(u),
    \label{eqn:general_nln}
\end{equation}
for some $F$,
with time forward map $f^t(u) = u + \int_0^t F(u)\dd t'$.

The (one parameter family of) Koopman operator(s) acts on functionals, or `observables', $\boldsymbol g$
of the state $u$ by shifting them along a trajectory of (\ref{eqn:general_nln}),
\begin{equation}
    \mathscr K^t \boldsymbol g(u) := \boldsymbol g(f^t(u)).
\end{equation}
This perspective is useful due to the linearity of the Koopman operator.
In particular, the eigenfunctions of $\mathscr K^t$ are scalar observables with an exponential dependence on time,
\begin{equation}
    \mathscr K^t \varphi_{\lambda}(u) = \varphi_{\lambda}(f^t(u)) := \varphi_{\lambda}(u)e^{\lambda t},
\end{equation}
and therefore constitute a coordinate system for representing arbitrary observables in which the nonlinear evolution \emph{appears} to be linear, 
\begin{equation}
    \label{eqn:kmd}
    \mathscr K^t \boldsymbol g(u) = \boldsymbol g(f^t(u))=\sum_{n=0}^{\infty} \varphi_{\lambda_n}(u)e^{\lambda_n t}\hat{\boldsymbol g}_n,
\end{equation}
where $\hat{\boldsymbol g}_n$ are Koopman modes for the observable $\boldsymbol g$.

Often the desire is to find a representation like (\ref{eqn:kmd}) for the function describing the state itself, $u$, so that for equation (\ref{eqn:general_nln}),
\begin{equation}
    u(x) = \sum_{n=0}^{\infty}\varphi_{\lambda_n}(u) \hat{u}_n(x).
    \label{eqn:u_decomp}
\end{equation}
In this notation, $u$ is viewed as a family of observables \emph{parameterised} by $x$.

The recent work by \citet{Page2019} demonstrated that separate Koopman decompositions (\ref{eqn:u_decomp}) can be constructed around simple invariant solutions of (\ref{eqn:general_nln}), and in general multiple decompositions will be required for a given trajectory as it wanders between unstable exact solutions.
In this work our focus is on spatially localised dynamics, which typically require multiple Koopman decompositions in both time \emph{and} space to represent the full nonlinear evolution.

\subsection{Motivating example: a front in the Burgers equation}
\label{sec:burgers}
The Burgers equation was considered by \citet{Page2018}, who used the Cole-Hopf transformation to derive a Koopman decomposition for the state variable $u$.
In that study, only trajectories running down to the trivial solution were considered.
Here, our focus is on travelling waves.
The Burgers equation is defined by,
\begin{equation}
    F(u):= - u\partial_x u + \nu\partial^2_xu,
\end{equation}
and supports a variety of equilibria and travelling wave solutions \citep{Benton1972}.
We consider a right-propagating front with $u(x\to -\infty)=U_{\infty}$ and $u(x\to\infty)=0$,
\begin{equation}
    u(x,t) = c\left[1 - \text{tanh}\left(\frac{c}{2\nu}\left(x-ct\right)\right)\right],
\end{equation}
where the propagation speed $c:=U_{\infty}/2$.

In the approach of \citet{Page2018}, Koopman eigenfunctions for the Burgers equation were obtained by exploiting the Cole-Hopf transformation and performing a Koopman mode decomposition (KMD) of the linearising variable.
A KMD for the velocity field was then found by inverting this transformation.
While such an approach should also be possible here, we instead derive the KMD(s) for the propagating front via a Laplace transform approach \citep{Page2019}.
This approach is more appropriate here, as it identifies regions in the $x-t$ plane where a particular KMD is convergent.

In \citet{Page2018} it was shown that the Koopman eigenvalues of the Burgers equation are all real.
Therefore, we adopt the following ansatz for the velocity field:
\begin{equation}
    u(x,t) = \int_{-\infty}^{\infty}v(-\lambda;x)\varphi_{-\lambda}(u) e^{-\lambda t}\dd \lambda,
    \label{eqn:burgers_ansatz}
\end{equation}
where $v(\lambda;x)$ is a Koopman mode \emph{density} for the observable $u$, which is parameterised by $x$.

Equation (\ref{eqn:burgers_ansatz}) is a bilateral Laplace transform with time as the transform variable.
The Koopman mode density can be obtained by inverting the transform by integration along a Bromwich contour in the complex-$t$ plane,
\begin{align}
    v(-\lambda;x)\varphi_{-\lambda}(u) &= \frac{1}{2\pi i}\int_{\gamma - i \infty}^{\gamma + i \infty}u(x,t)e^{\lambda t} \dd t \nonumber \\
                                      &= \frac{c}{\pi i}\int_{\gamma - i \infty}^{\gamma + i \infty}\frac{e^{\lambda t}}{1 + \exp{\left[\frac{c}{\nu}(x-ct)\right]}} \dd t.
                                      \label{eqn:burg_inv}
\end{align}
For unilateral Laplace transforms, convergence is assured by selecting $\gamma$ to lie to the right of the singularities of the integrand.
For the bilateral transform, $\gamma$ can be selected to the right or left of the singularities
(the contour then closed to the left or right respectively)
provided that the Koopman mode density vanishes below or above a critical value of $\lambda$ respectively \citep{Page2019}.
This results in two possible Koopman mode densities.
In practice, one is associated with exponentially decaying Koopman eigenvalues, the other with exponential growth.

The inversion integrand (\ref{eqn:burg_inv}) has simple poles at $t_n = x/c +i \pi(2n+1)\nu/c^2$, $n \in \mathbb Z$.
The inversion can therefore be accomplished by selecting either $\gamma > x/c$ and closing to right or $\gamma < x/c$ and closing to the left,
a choice which yields a convergent KMD either upstream ($x<ct$) or downstream ($x > ct$) of the front.
The solution procedure is almost identical for both cases, and we discuss only the upstream calculation in detail.

For the upstream expansion, $\gamma > x/c$, we close the contour in a large semicircle to the \emph{left}.
The contribution to the integral from the semicircular contour vanishes for $\lambda> -c^2/\nu$, hence the corresponding Koopman mode density has support for $\lambda \in (-c^2/\nu,\infty)$ and the upstream KMD is
\begin{equation}
    u(x,t) = \int_{-c^2/\nu}^{\infty} v_-(-\lambda;x)\varphi_{-\lambda}(u)e^{-\lambda t}\dd \lambda,
    \label{eqn:b_upstream_int}
\end{equation}
where
\begin{align}
    \begin{split}
    &v_-(-\lambda;x)\varphi_{-\lambda}(u)
    \\&\qquad= \frac{c}{\pi i}\oint_C\frac{e^{\lambda t}}{1 + \exp{\left[\frac{c}{\nu}(x-ct)\right]}} \dd t \\
    &\qquad= 2c\sum_{n=-\infty}^{\infty} \text{Res}\left(\frac{e^{\lambda t}}{1 + \exp{\left[\frac{c}{\nu}(x-ct)\right]}}, t_n\right),
    \end{split}
\end{align}
where $C$ is the closed contour built from the Bromwich contour and the large semicircle.
Evaluating the residues at the poles, we find
\begin{align}
    \begin{split}
    &v_-(-\lambda;x)\varphi_{-\lambda}(u)
    \\&\qquad= 2c \sum_{n=-\infty}^{\infty} \frac{\nu}{c^2} \exp{\left[\lambda\left(\frac{x}{c} +i \pi(2n+1)\frac{\nu}{c^2}\right)\right]}
    \\&\qquad= \frac{2\nu}{c} (-1)^{\lambda \nu/c^2}\exp{\left(\frac{ \lambda x}{c}\right)} \sum_{k=-\infty}^{\infty} \delta\left(k-\frac{\lambda\nu}{c^2}\right),
    \end{split}
\end{align}
using the identity for generalised functions $\sum_n e^{2\pi i n t} = \sum_k \delta(k-t)$.
Inserting the upstream density in (\ref{eqn:b_upstream_int}) yields the upstream KMD,
\begin{equation}
    u(x,t) = 2c \sum_{k=0}^\infty (-1)^k\exp{\left[\frac{kc}{\nu}(x-ct)\right]} ,
    \label{eqn:bur_up}
\end{equation}
valid for $x<ct$.

\begin{figure}
    \centering
    \includegraphics[width=\columnwidth]{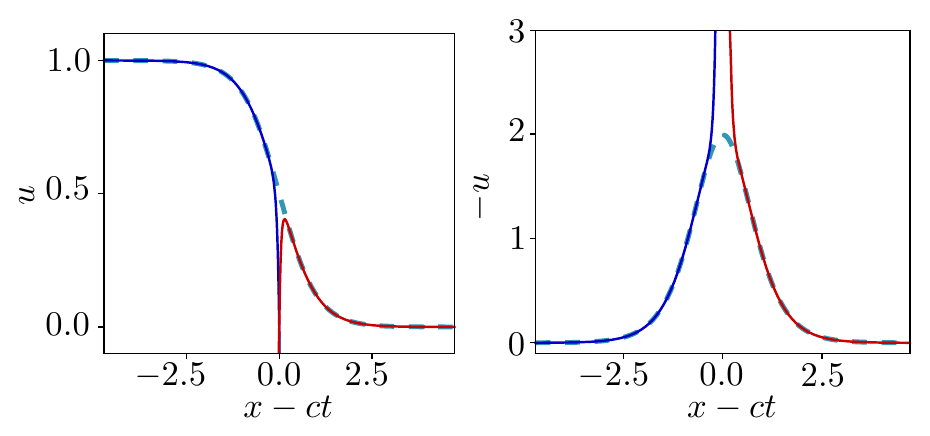}
    \caption{Simple travelling wave solutions to the Burgers (left) and KdV (right) equations visualised in a co-moving frame along with the respective upstream (blue) and downstream (red) Koopman expansions. Series are truncated at $N=10$ in all cases.}
    \label{fig:front}
\end{figure}

A similar approach with $\gamma < x/c$ yields
\begin{align}
    \begin{split}
    &v_+(-\lambda;x)\varphi_{-\lambda}(u)
    \\ &\qquad = -\frac{2\nu}{c} (-1)^{\lambda \nu/c^2}\exp{\left(\frac{ \lambda x}{c}\right)} \sum_{k=-\infty}^{\infty} \delta\left(k-\frac{\lambda\nu}{c^2}\right),
    \end{split}
\end{align}
with the KMD for the velocity downstream
\begin{align}
    u(x,t) &= \int_{-\infty}^{0}v_+(-\lambda;x)\varphi_{-\lambda}(u)e^{-\lambda t}\dd \lambda \nonumber \\
    &= -2c \sum_{k=1}^{\infty}(-1)^k \text{exp}\left[-\frac{kc}{\nu}\left(x-ct\right)\right],
    \label{eqn:bur_down}
\end{align}
valid for $x>ct$.
Both the downstream expansion (\ref{eqn:bur_down}) and the upstream expansion (\ref{eqn:bur_up}), truncated at $N=10$ terms, are overlayed onto the true travelling front solution in figure \ref{fig:front}.
The loss of convergence in both expansions at $x-ct=0$ is clear.

There is a simple dynamical systems interpretation to these results:
Under the ansatz of travelling-wave dynamics, the Burgers equation reduces to a simple one-dimensional (nonlinear) ordinary differential equation and the front depicted in figure \ref{fig:front} is a heteroclinic connection between the (unstable) trivial solution at the origin and the (stable) equilibrium $u=U_{\infty}$.
The pair of Koopman decompositions found above thus corresponds to expansions about these two equilibria, which both breakdown at the same ``crossover point'' in state space \citep[see also][]{Page2019}.
These equilibria have one-dimensional linear subspaces, and the associated Koopman decompositions begin with eigenvalues corresponding to these unstable/stable linear dynamics, $\mp c^2/\nu$.
The higher order terms in the expansion then correspond to integer powers of the associated Koopman eigenfunction.

\section{Koopman decomposition of Korteweg-de-Vries equation}
\label{sec:kdv}

The Korteweg-de-Vries (KdV) equation is the canonical and simplest example of a nonlinear dispersive wave equation.
It is defined by
\begin{equation}
    \label{eq:kdv}
    F(u) := -\partial^3_x u + 6 u\partial_x u.
\end{equation}
The term $\partial^3_x u$ makes this a dispersive wave equation, and $u\partial_x u$
is a nonlinear self-advection term. 
Equation (\ref{eq:kdv}) naturally arises as the inclusion of simple nonlinearity in a number of wave phenomena,
including internal waves in a stratified fluid.

In an early example of the numerical solution of PDEs, \citet{Zabusky1965} simulated
the KdV equation and discovered the rich behaviour of so-called `solitons'. These exact coherent structures
of the PDE are strongly stable. They can interact with one another and preserve their form post-interaction.
The behaviour of solitons led to the development of the inverse scattering transform (IST),
which can be used to analytically solve KdV as well as a number of other, more complicated, so-called
`integrable' equations.

\subsection{Single-soliton solution}
\label{sec:kdvintro}
The canonical one-soliton solution to KdV is given by
\begin{equation}
    \label{eq:1sol}
    u(x,t) = -2 \sech^2 \left( x-4t \right),
\end{equation}
which is a simple travelling wave propagating to the right.
Note that $u<0$, which is the case for all soliton solutions of (\ref{eq:kdv}). 

We will follow the methodology outlined for the front in the Burgers equation in \S \ref{sec:burgers} and assume
that the Koopman eigenvalues required to described the evolution of (\ref{eq:1sol}) are real.
This assumption will be justified \S \ref{sec:kdveigenpairs}, where we derive the Koopman eigenfunctions required to describe arbitrary soliton evolutions.

Expressing the evolution as an integral over a Koopman mode \emph{density} (see \S \ref{sec:burgers}), $v(\lambda;x)$,
\begin{equation}
    \label{eq:nonspecifickmd}
    -2 \sech^2 \left( x-4t \right) = \int_{-\infty}^{\infty} v(-\lambda;x)\varphi_{-\lambda}(u) e^{-\lambda t} \dd \lambda.
\end{equation}
This Laplace transform (transform variable $t$) can be inverted in the normal way to give
\begin{align}
    \label{eq:inverselaplace}
    \begin{split}
    v(-\lambda;x)\varphi_{-\lambda}(u) &= \frac{1}{2\pi i} \int_{\gamma-i\infty}^{\gamma+i\infty} -2 \sech^2 \left( x-4t \right) e^{\lambda t} \dd t
    \\
    &= -\frac{1}{\pi i} \int_{x-4\gamma-i\infty}^{x-4\gamma+i\infty} \frac{e^{\lambda (x -\xi)/4}}{\left(e^\xi+e^{-\xi}\right)^2} \dd \xi,
    \end{split}
\end{align}
where $\xi := x-4t$.
Similar to the Burgers equation example presented in \S \ref{sec:burgers}, we can close the contour for this integral in two different directions,
yielding a pair Koopman decompositions which hold upstream/downstream of the soliton.

Closing the contour to the left, we label the Koopman modes as $v_+$, with $v_+(\lambda;x)=0$ for $\lambda>2$.
Then (\ref{eq:nonspecifickmd}) becomes
\begin{equation}
    \label{eq:kmdplus}
    -2 \sech^2 \left( x-4t \right) = \int_{-\infty}^{2} v_+(-\lambda;x)\varphi_{-\lambda}(u) e^{-\lambda t} \dd \lambda.
\end{equation}

Equation (\ref{eq:inverselaplace}) has second order poles at $\xi_n = i \pi(2n+1)/2$, $n \in \mathbb Z$. 
The residue theorem gives, for $\lambda<2$,
\begin{align}
    \begin{split}
    &v_+(-\lambda;x)\varphi_{-\lambda}(u) \\
    &\qquad= -2\sum_{n=-\infty}^\infty \text{Res}\left(\frac{e^{\lambda (x -\xi)/4}}{\left(e^\xi+e^{-\xi}\right)^2}, i \pi(2n+1)/2\right)
    \\
    &\qquad= -\lambda e^{\lambda x /4} e^{-i\pi \lambda/8}\sum_{k=-\infty}^{\infty} \delta\left(8k-\lambda\right).
    \end{split}
\end{align}

Substituting this into (\ref{eq:kmdplus}), we find a decomposition
\begin{equation}
    \label{eqn:upstreamsol}
    -2 \sech^2 \left( x-4t \right)
    =\sum_{k=1}^{\infty}  8k (-1)^k e^{-2k x}e^{8k t}.
\end{equation}
This expansion involves Koopman eigenvalues $\{8k:k\in \mathbb{N}\}$, with corresponding Koopman modes
$e^{-2kx}$.
In this derivation, it is not possible to determine the Koopman eigenfunctions $\{\varphi_{\lambda}(u)\}$ in their general form.

Equation (\ref{eqn:upstreamsol}) is a convergent expansion for $x>4t$, i.e. downstream of the peak of the soliton.
Analogous behaviour was seen in the front solution to Burgers equation (e.g. (\ref{eqn:bur_up})), which suggests that the
need for multiple Koopman decompositions to describe nonlinear wave evolution is generic.
The upstream expansion for the one soliton solution to KdV can be obtained by closing the contour to the right, which yields
\begin{equation}
    \label{eqn:downstreamsol}
    2 \sech^2 \left( x-4t \right)= \sum_{k=1}^{\infty} 8k (-1)^ke^{2kx} e^{-8k t},
\end{equation}
which could also be anticipated from symmetry.
The upstream expansion is convergent for $x<4t$ and involves Koopman eigenvalues $\{-8k:k\in \mathbb{N}\}$ -- temporally decaying modes.

Similar to the Burgers equation, there is a simple dynamical systems interpretation to these results which
rests on the fact that Koopman expansions appear to be defined about simple invariant solutions of the governing equation, and connecting orbits between such solutions contain a crossover point where one expansion fails and another takes over.
The one soliton solutions to the KdV equation in a travelling frame, $\xi :=x-ct$ (here $c=4$),
are homoclinic orbits about the trivial solution $u=0$, and the crossover point $\xi=0$ divides these trajectories into `repelling' and `attracting' sections.
The Koopman expansions for these sections of the orbit are built from eigenfunctions which are integer powers of the Koopman eigenfunctions associated with the linear subspace around $u(\xi)=0$ and have eigenvalues $\pm \sqrt{c}$ (i.e. $\pm c^{3/2}$ in the lab frame).

These effects have interesting consequences for describing more complex dynamics -- soliton interactions -- in terms of Koopman expansions.
In order to generalise the approach above, we will use the inverse scattering transform \citep[e.g.][]{DrazinSoliton} to derive Koopman eigenfunctions for the KdV equation in their general form, which will allow us to examine these more interesting situations.

\subsection{Inverse scattering method}
\label{sec:inversescattering}
The inverse scattering method is one the most celebrated results of twentieth century mathematics.
It can be used to solve a variety of nonlinear PDEs, including the nonlinear Schr\"odinger equation and the
sine-Gordon equation \citep{Ablowitz1974}.
In the inverse scattering approach, the solution to the nonlinear PDE, $u(x,t)$, is treated as a
potential in a linear scattering problem in which time appears parametrically.
It can be shown that the scattering data (the eigensolutions of the scattering problem) evolve
\emph{linearly} as $u(x,t)$ evolves according to its nonlinear evolution equation.
Therefore, the scattering data can be obtained for all time from the initial condition $u(x,0)$ alone.
The solution to the nonlinear PDE at any time can then be extracted from the scattering data via an
inverse scattering transform, which amounts to the solution of a linear integral equation.
The existence of a linearising transform allows us to derive Koopman eigenfunctions, which can then be used to construct Koopman decompositions for the state variable itself.

Here we concentrate on the specific case of KdV, for which the inverse scattering method
was first developed \citep{Gardner1967}.
Throughout, we follow the notation and conventions of \citet{DrazinSoliton}. 
Let $u_0(x)$ be some initial condition for the KdV equation on the real line, with $u_0(x)\to 0$ as $x\to\pm\infty$.
The time evolution can then be obtained as follows:
\begin{enumerate}
    \item Solve the eigenvalue Sturm-Liouville scattering problem $\psi_{xx}+(\lambda-u_0)\psi=0$.
          The eigenvalue spectrum has a discrete negative part $\lambda=-\kappa_n^2$ for $n=1,2,\ldots,N$,
          and a continuous positive part $\lambda=k^2$. The eigenvalues and their corresponding
          eigenfunctions are called the `scattering data'.
    \item It is then possible to predict how the scattering data will evolve as $u$ evolves from
          $u_0$ according to the KdV equation. In particular, it is sufficient to consider the
          `reflection coefficient' $b(k)$ for the continuous spectrum and $\{c_n\}$ for the discrete spectrum.
          These are defined by requiring that the eigenfunctions $\psi\sim e^{-ikx}+b(k)e^{ikx}$ or
          $\psi\sim c_n e^{-\kappa_n x}$ as $x\to\infty$. The latter (discrete) case is normalised
          so that $\int_{-\infty}^\infty \psi^2 \dd x = 1$.

          The scattering data evolve according to the linear equations
          \begin{subequations}
              \begin{align}
                  \frac{\dd b}{\dd t}&=8ik^3b, \\
                  \frac{\dd c_n}{\dd t}&=4\kappa_n^3 c_n,
                  \label{eqn:scat_evol}
              \end{align}
        \end{subequations}
        as the potential $u(x)$ evolves according to the KdV equation.

    \item Given the scattering data at initial time, one can then calculate $u(x,t)$ at
          some future time $t$ through
          `inverse scattering', which amounts to solving the Marchenko equation,
          \begin{align}
            \label{eq:marchenko}
            \begin{split}
            &K(x,z,t)+F(x+z,t)\\&+\int_x^\infty K(x,y,t)F(y+z,t)\dd y=0,
            \end{split}
          \end{align}
          for $K$, where
          \begin{align*}
            F(x,t)=&\sum_{n=1}^N c_n^2 \exp{(8\kappa_n^3 t -\kappa_n x)}
            \\&+\frac{1}{2\pi}
          \int_{-\infty}^\infty b(k) \exp{(8ik^3t+ikx)}\dd x.
        \end{align*}
          In all but the simplest cases, this must be done numerically.
          The velocity is then obtained via $u(x,t)=-2\left(\partial_x K(x,z,t)|_{z=x}+\partial_z K(x,z,t)|_{z=x}\right)$.
\end{enumerate}

\subsection{Koopman eigenpairs of the KdV equation}
\label{sec:kdveigenpairs}
With the inverse scattering transform in mind, we now define a family of observables $c_\kappa(u)$,
where $\kappa$ is a positive real number,
on the state space for the unbounded KdV equation. The value of $c_\kappa(u)$, a real number,
can be computed as follows:
First, determine whether the ordinary differential equation $\psi_{xx} - (\kappa^2+u(x))\psi = 0$
    has a non-trivial, square-integrable solution, with $\psi$ decaying exponentially as $x\to\pm\infty$.
 If it does, the solution is made unique by requiring $\int_{-\infty}^\infty \psi^2 \dd x=1$.
 In the limit $x\to \infty$, $\psi\sim A e^{-\kappa x}$ for some $A$, which allows us to define $c_\kappa(u)=A$.
 If there is no solution to the Sturm-Liouville problem, define $c_\kappa(u)=0$.

Due to their linear evolution equations (\ref{eqn:scat_evol}), it is clear that the scattering data are Koopman eigenfunctions of the nonlinear KdV equation,
\begin{equation}
    \mathscr{K}^t c_\kappa(u) = c_{\kappa}(f^t(u)) =  e^{4\kappa^3t}c_\kappa(u),
\end{equation}
i.e. $c_{\kappa}(u) = \varphi_{\lambda_{\kappa}}(u)$, the Koopman eigenfunction with Koopman eigenvalue $\lambda_{\kappa}=4\kappa^3$.

We note that the same approach can be used to construct a family of Koopman eigenfunctions with purely imaginary Koopman eigenvalues from the
reflection coefficients $b(k)$ associated with the continuous spectrum of the scattering problem.
Because of difficulties solving the integral equation in cases where $b(k)\neq 0$, we consider only
`reflectionless potentials' where $b(k)\equiv0$.

Since the scattering data are sufficient to reconstruct the whole solution to the KdV equation,
we therefore assume that these Koopman eigenpairs, and their products, as discussed below, are sufficient
to find decompositions.

\subsection{Single-soliton revisited}
\label{sec:1sol}
Before examining soliton interactions, we will first revisit the
one soliton solution of the KdV equation considered in \S \ref{sec:kdvintro},
\begin{equation}
    u(x,0) = -2\sech^2 x,
\end{equation}
and use knowledge of the Koopman eigenfunctions and the inverse scattering approach to construct the Koopman decompositions.
From our family of Koopman eigenfunctions $c_\kappa$, only $c_1(u)$ is non-zero in this case, with $c_1(u_0)=\sqrt{2}$, and
there is no continuous spectrum in the scattering problem.
However, note that $c_\kappa$ can be raised to any power $a$ to give a Koopman eigenfunction with Koopman eigenvalue
$4a\kappa^3$ \citep{Mezic2013}.

Initially, we introduce as an ansatz a Koopman decomposition using only positive integer powers of
$c_1(u)$ -- i.e. one associated with exponential growth in time.
We will see that this approach yields the upstream expansion (\ref{eqn:upstreamsol}) found via the Laplace transform approach in \S \ref{sec:kdvintro}.
Rather than seeking a decomposition for $u(x)$ directly, we first decompose $K(x,z)$, the solution to the Marchenko equation described in \S\ref{sec:inversescattering}.
With our ansatz, we write
\begin{equation}
    K(u;x,z) = \sum_{n=1}^\infty{\hat{K}_n(x,z)c_1^n(u)} = \sum_{n=1}^\infty{\hat{K}_n(x,z)c_1^n(u_0)e^{4nt}},
\end{equation}
where $c_1^n(u_0) = 2^{n/2}$.
Note the change in notation to reflect that $K$ is an observable of the state, $u$, parameterised by $x$ and $z$.
The Marchenko equation (\ref{eq:marchenko}) now reads
\begin{align*}
    &\sum_{n=1}^\infty{\hat{K}_n(x,z)c^n_1(u_0)e^{4nt}} + 2 e^{8t-x-z}
    \\&+ \int_x^\infty \sum_{n=1}^\infty{\hat{K}_n(x,y)c_1^n(u_0)e^{4nt}} 2 e^{8t-y-z}\dd y = 0.
\end{align*}
Examining the $z$ dependence of the terms, it is apparent that $\hat{K}_n(x,z)=\hat{L}_n(x)e^{-z}$
for some $\hat{L}_n(x)$. We can therefore perform the integration, to give
\begin{align*}
    &\sum_{n=1}^\infty{\hat{L}_n(x) c_1^n(u_0)e^{4nt}} + 2 e^{8t-x}
    \\&+ \sum_{n=1}^\infty{\hat{L}_n(x) c_1^n(u_0) e^{(8+4n)t-2x}} = 0.
\end{align*}
Comparing coefficients of $e^{4pt}$, we have
\begin{align*}
    &\hat{L}_p(x)c_1^p(u_0) + \hat{L}_{p-2}(x)c_1^{p-2}(u_0)e^{-2x} \\&\quad=
    \begin{cases}
        -2e^{-x}, &p=2,\\
        0, &\mathrm{otherwise}.
    \end{cases}
\end{align*}
Assuming that the Koopman modes associated with the exponentially decaying eigenfunctions not included in the ansatz ($c_1^{-n}(u_0)$) are zero, $\hat{L}_n(x) = 0$ for $n<0$, this recurrence may be solved directly to give
\begin{equation}
    \hat{L}_n(x) = \begin{cases}
        0, &n\;\mathrm{odd},\\
        (-1)^{n/2} 2^{1-n/2} e^{-(n-1)x}, &n\;\mathrm{even}.
    \end{cases}
\end{equation}
The resulting Koopman decomposition for $K$ is then
\begin{equation}
    K(u;x,z) = \sum_{n=1}^\infty (-1)^{n} 2^{1-n}e^{-(2n-1)x-z} c_1^{2n}(u_0) e^{8nt},
\end{equation}
and a Koopman decomposition for $u$ can be obtained from $u = -2\left(\partial_x K(u;x,z)|_{z=x}+\partial_z K(u;x,z)|_{z=x}\right)$, giving
\begin{align}
    u(x,t) &= \sum_{n=1}^{\infty}(-1)^n 2^{3-n}e^{-2nx} c_1^{2n}(u_0)e^{8nt}, \nonumber \\
    &= \sum_{n=1}^\infty (-1)^{n} 8n e^{-2n(x-4t)}.
    \label{eqn:upstream_kmd_ck}
\end{align}
This is a Koopman decomposition, using Koopman eigenfunctions $c_1^{2n}(u)$ with Koopman eigenvalues
$8n$, and Koopman modes $\hat{u}_{2n}(x) = 8n (-1/2)^{n} e^{-2nx}$.

Equation (\ref{eqn:upstream_kmd_ck}) matches that found in \S \ref{sec:kdvintro} using the inverse Laplace transform (\ref{eqn:upstreamsol}).
To find the second Koopman expansion, valid downstream of the soliton, 
we would begin with the ansatz,
\begin{equation}
    K(u;x,z) = \sum_{n=1}^\infty{\hat{K}_n(x,z)c_1^{-n}(u)},
\end{equation}
i.e. an expansion in exponentially decaying Koopman eigenfunctions.

In summary,
we have used the inverse scattering transform approach to identify Koopman eigenfunctions and eigenvalues of the KdV equation
and shown how different sets of eigenfunctions are required in different regions of space-time to express localised nonlinear
wave evolution in the form of a Koopman decomposition.
We now extend this approach to examine more complex dynamics involving soliton interactions,
where the number of possible Koopman decompositions increases dramatically.
Selecting the appropriate decomposition for a given region of the $x-t$ plane depends on the relative positions of all solitons.

\subsection{Multiple solitons}
\label{sec:2sol}
The method presented in \S\ref{sec:1sol} can be generalised to an arbitrary but finite
number of solitons, so long as the initial condition has no continuous spectrum in the scattering problem.
To demonstrate the approach, we examine in detail the interaction of two solitons.

With two solitons, we now have two non-zero scattering eigenvalues $\kappa_1$ and $\kappa_2$,
with corresponding Koopman eigenfunctions $c_{\kappa_1}(u)$ and $c_{\kappa_2}(u)$ and Koopman eigenvalues
$4\kappa_1^3$ and $4\kappa_2^3$.
The eigenfunctions $c_{\kappa_1}(u)$ and $c_{\kappa_1}(u)$ can be raised to arbitrary
powers to produce further Koopman eigenfunctions, but we can now also multiply them \citep{Mezic2013}.
As was found in the one-soliton case, only even powers are required, since $c_\kappa^2$ rather than $c_\kappa$ appears in the Marchenko equation. 
The possible combinations of $c_{\kappa_1}(u)$ and $c_{\kappa_2}(u)$ thus yield a set of Koopman eigenfunctions of the form
\begin{equation*}
     c_{\kappa_1}^{2j}(u) c_{\kappa_1}^{2k}(u), \qquad (j,k)\in\mathbb{Z}^2,
\end{equation*}
with corresponding Koopman eigenvalues $4\kappa_1^3\cdot2j + 4\kappa_2^3\cdot 2k = 8(\kappa_1^3 j + \kappa_2^3 k)$.
If $\kappa_1$ and $\kappa_2$ are both rational numbers then the Koopman eigenvalues will be degenerate, an effect that has also been observed
in Koopman decompositions of the Burgers equation \citep{Page2018}.

With two scattering eigenvalues, the Marchenko equation (\ref{eq:marchenko}) becomes
\begin{align}
    \begin{split}
    &K(x,z,t)
    \\&+c_{\kappa_1}^2 \exp{(8\kappa_1^3 t -\kappa_1 (x+z))}
    \\&+c_{\kappa_2}^2 \exp{(8\kappa_2^3 t -\kappa_2 (x+z))}
    \\&+\int_x^\infty K(x,y,t)c_{\kappa_1}^2 \exp{(8\kappa_1^3 t -\kappa_1 (y+z))}\dd y
    \\&+\int_x^\infty K(x,y,t)c_{\kappa_2}^2 \exp{(8\kappa_2^3 t -\kappa_2 (y+z))}\dd y=0.
    \end{split}
    \label{eqn:2sol_march}
  \end{align}
The $z$-dependence of the terms in (\ref{eqn:2sol_march}) implies $K(x,z,t)$ is of the form
\begin{equation}
    K(x,z,t)=L^{(1)}(x,t)e^{-\kappa_1 z}+L^{(2)}(x,t)e^{-\kappa_2 z},
    \label{eqn:2sol_K_L}
\end{equation}
which reduces (\ref{eqn:2sol_march}) to a pair of coupled equations:
\begin{align}
    \label{eq:marchenkoL}
    \begin{split}
    &L^{(1)}(x,t)+c_{\kappa_1}^2 e^{8\kappa_1^3 t -\kappa_1 x}
    \\&+\frac{1}{2\kappa_1}L^{(1)}(x,t)c_{\kappa_1}^2e^{8\kappa_1^3t-2\kappa_1x}
    \\&+\frac{1}{\kappa_1+\kappa_2}L^{(2)}(x,t)c_{\kappa_1}^2e^{8\kappa_1^3t-(\kappa_1+\kappa_2)x} = 0,
    \end{split}
    \\
    \begin{split}
        &L^{(2)}(x,t)+c_{\kappa_2}^2 e^{8\kappa_2^3 t -\kappa_2 x}
        \\&+\frac{1}{\kappa_1+\kappa_2}L^{(1)}(x,t)c_{\kappa_1}^2e^{8\kappa_1^3t-(\kappa_1+\kappa_2)x}
        \\&+\frac{1}{2\kappa_2}L^{(2)}(x,t)c_{\kappa_1}^2e^{8\kappa_1^3t-2\kappa_2x} = 0.
    \end{split}
\end{align}

We propose Koopman decompositions for the observables $L^{(1)}$ and $L^{(2)}$ of the form
\begin{multline}
    \label{eq:2solansatz}
    L^{(1,2)}(u;x) = \\\sum_{j}\sum_{k} \hat{L}^{(1,2)}_{j,k}(x,z) c_{\kappa_1}^{2j}(u_0) c_{\kappa_1}^{2k}(u_0) e^{8(\kappa_1^3 j + \kappa_2^3 k)t}.
\end{multline}
As found in the single soliton case, the range of values over which we sum $j$ and $k$, or equivalently whether the expansion is constructed using exponentially growing or decaying modes (or a combination),
implicitly selects a region of space-time in which the expansion converges.
The various choices are discussed in detail below.
Substituting the ansatz (\ref{eq:2solansatz}) into (\ref{eq:marchenkoL})
and comparing coefficients of exponentials (assuming no degeneracy) yields the recurrence relations
\begin{align}
    \begin{split}
    \label{eq:2sol1}
    \hat{L}^{(1)}_{j,k} + \frac{1}{2\kappa_1} \hat{L}^{(1)}_{j-1,k}e^{-2\kappa_1 x} + \frac{1}{\kappa_1+\kappa_2} \hat{L}^{(2)}_{j-1,k}e^{-(\kappa_1+\kappa_2) x}  \\
    =
    \begin{cases}
        -e^{-\kappa_1 x},&j=1,k=0,\\
        0,&\mathrm{otherwise},
    \end{cases}
    \end{split}
    \\
    \begin{split}
        \label{eq:2sol2}
        \hat{L}^{(2)}_{j,k} + \frac{1}{\kappa_1+\kappa_2} \hat{L}^{(1)}_{j,k-1}e^{-(\kappa_1+\kappa_2) x} + \frac{1}{2\kappa_2} \hat{L}^{(2)}_{j-1,k}e^{-2\kappa_2 x}  \\
        =
        \begin{cases}
            -e^{-\kappa_2 x},&j=0,k=1,\\
            0,&\mathrm{otherwise}.
        \end{cases}
    \end{split}
\end{align}
With some rearrangement, these can be solved straightforwardly for $j$ and $k$ either increasing or decreasing,
and various boundary conditions are therefore possible. The solutions are too complicated to include here,
but can be found using a computer algebra system. 
As described previously in the one soliton calculation, the Koopman decomposition for the pair of observables $L^{(1,2)}(u;x)$
can be converted into a Koopman decompositions for $K(u;x,z)$ via equation (\ref{eqn:2sol_K_L}), before the decomposition
for the velocity is obtained from $u(x,t)=-2\left(\partial_x K(x,z,t)|_{z=x}+\partial_z K(x,z,t)|_{z=x}\right)$ \citep{DrazinSoliton}.
We now discuss the solutions for various boundary conditions in the above recurrence relations.

\begin{figure}
\includegraphics[width=\columnwidth]{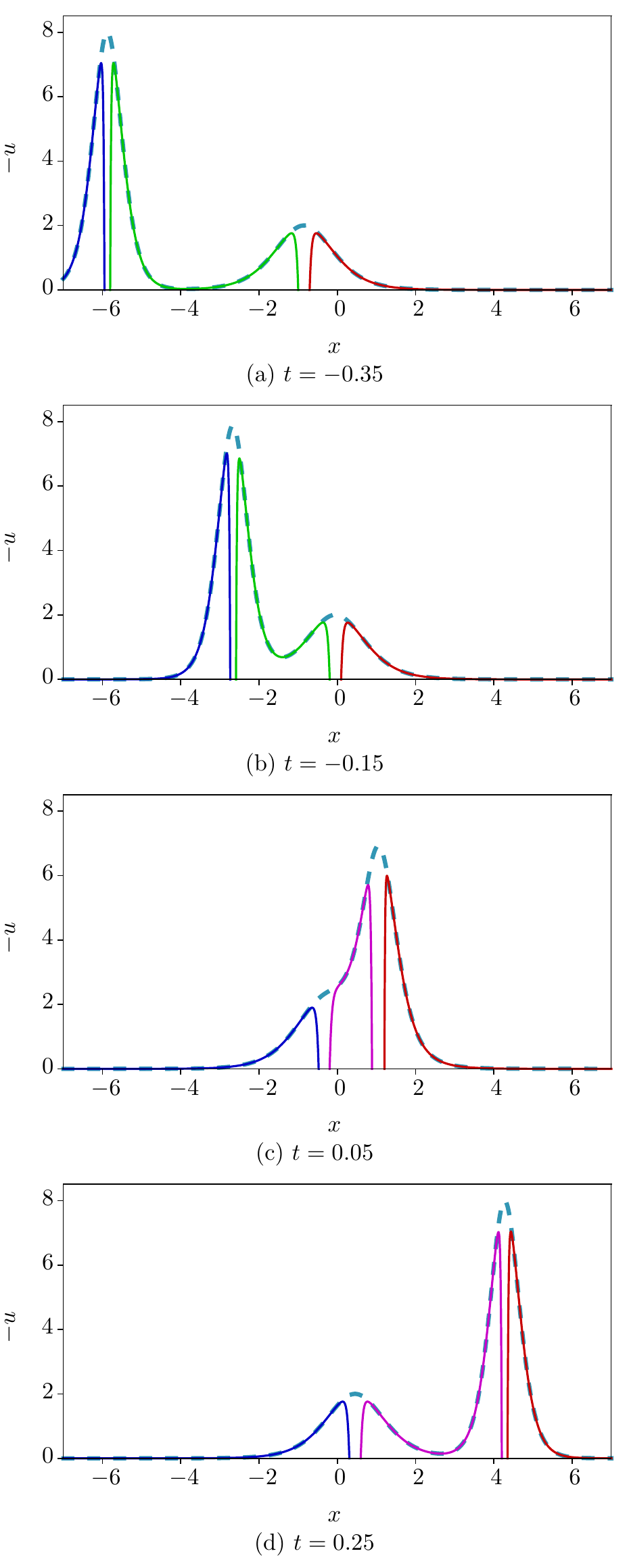}
\caption{Truncated Koopman decompositions for the 2-soliton solution (\ref{eq:2solanalytic}) (shown as dashed line), at different times.
For $t<0$, the decomposition with Koopman eigenfunctions $c_1^j c_2^k$ with $j\leq0$ and $k\geq0$ (in green) must be used between the solitons, whereas $k\geq0$ and $j\leq0$ (in pink) does not converge, and is completely
off the scale of the plot. The reverse is true for $t>0$. The $j\leq0$, $k\leq0$ expansion (blue) and $j\geq0$, $k\geq0$ (red) are needed at all times, upstream and downstream, respectively, of both solitons.}
\label{fig:2sol}
\end{figure}

First, we adopt the assumption that
$\hat{L}^{(1)}_{j,k}$ and $\hat{L}^{(2)}_{j,k}$ are zero for $j<0$ and $k<0$,
or equivalently seek to build a solution using only temporally growing modes.
In this instance,
an expansion is obtained that is valid downstream of both solitons.
The velocity field resulting from this solution for $L^{(1,2)}$ is reported in figure \ref{fig:2sol} (the red curves) for a particular choice of
$\kappa_1$ and $\kappa_2$ which is discussed further below.

On the other hand, if both $\hat{L}^{(1)}_{j,k}$ and $\hat{L}^{(2)}_{j,k}$ are assumed to be zero for $j>0$ and $k>0$, an expansion is obtained
in terms of only temporally decaying modes.
As shown in figure \ref{fig:2sol} (blue curves), this expansion converges upstream of both solitons.
Both this time-decaying expansion and the temporally growing expansion discussed above are analogous to the expansions obtained for the single soliton (see \S \ref{sec:kdvintro} and \S \ref{sec:1sol}).
However, the inclusion of products of the Koopman eigenfunctions allows the `linear' Koopman decompositions to represent the dynamics upstream and
downstream of the solitons during their interaction.
As shown in figure \ref{fig:2sol}, these expansions apply both before and after the faster soliton overtakes the slower.

A more interesting ansatz is to assume $\hat{L}^{(1)}_{j,k}=0$ and $\hat{L}^{(2)}_{j,k}=0$ for $j<0$ but $k>0$.
This amounts to a decomposition involving growing modes associated with the $\kappa_1$ eigenvalue (i.e. those that describe the evolution upstream of soliton 1)
but decaying modes associated with the $\kappa_2$ eigenvalue (describing the evolution downstream of soliton 2).
An example of this expansion, which describes the evolution between the solitons up to (and including part of) their interaction, is shown in figure \ref{fig:2sol} (green curves).
The products in the Koopman expansion of the form $c_{\kappa_1}^j(u)c^k_{\kappa_2}(u)$ allow for a `linear' representation of the strongly nonlinear dynamics between the solitons as they interact.

However, as the faster soliton approaches the slower, the region of space in which this decomposition holds shrinks and eventually vanishes.
For a Koopman decomposition which holds between the solitons post-interaction, it is necessary to instead assume
$\hat{L}^{(1)}_{j,k}=0$ and $\hat{L}^{(2)}_{j,k}=0$ for $j>0$ and $k<0$, i.e. an ansatz using the unstable eigenvalues
for the $\kappa_2$ soliton and the stable eigenvalues associated with the $\kappa_1$ soliton.
This expansion is shown in pink in figure \ref{fig:2sol}.

The particular two soliton interaction reported in figure \ref{fig:2sol} is the `classical' two soliton solution
\citep[see e.g.][]{DrazinSoliton} defined by the initial condition,
\begin{equation}
    u(x,0) = -6\sech^2{x},
    \label{eqn:2sol_ic}
\end{equation}
for which the KdV equation has the known analytical solution,
\begin{equation}
    \label{eq:2solanalytic}
    u(x,t) = -12 \frac{3+4\cosh(2x-8t)+\cosh(4x-64t)}{\left(3\cosh(x-28t)+\cosh(3x-36t)\right)^2}.
\end{equation}
This solution is particularly useful when assessing the crossover between the multiple Koopman decompositions
owing to the fact that the initial condition (\ref{eqn:2sol_ic}) corresponds to the temporal ``midpoint'' in the interaction
between the two solitons which separate as $t\to \pm \infty$.
In fact, precisely when $t=0$, neither of the interior decompositions (the green and pink curves in figure \ref{fig:2sol}) are valid, and they are nowhere pointwise
convergent to a finite value (not shown).
When $t$ is very small, a very large number of terms is required for the expansions to well approximate the true solution near the solitons.

Another consequence of using the solution defined by (\ref{eq:2solanalytic}) is the occurrence of degeneracy in the Koopman eigenvalues.
The scattering problem for this potential gives discrete eigenvalues of $\kappa_1 = 1$ and $\kappa_2 = 2$.
These values correspond to Koopman eigenvalues $4\kappa_1^3 = 4$ and $4\kappa_2^3 = 32$ and
normalisation coefficients (Koopman eigenfunctions) $c_1(u_0)=\sqrt{6}$ and $c_2(u_0)=2\sqrt{3}$ respectively \citep{DrazinSoliton}.
The fact that the two Koopman eigenvalues are both proportional to perfect cubes, coupled with allowance for both exponentially decaying and
growing modes, causes the degeneracy.
For example, the combinations $(j,k) = (0,2)$ (eigenfunction $c_2^4(u)$) and $(j,k)=(8,1)$
(eigenfunction $c_1^{16}(u)c_2^2(u)$) both share the eigenvalue $128$.
In the degenerate case, the recurrence relations presented above (\ref{eq:2sol1}, \ref{eq:2sol2}) are now only one possible solution to the
Marchenko equation.
However, considering the nondegenerate situation with $\kappa_1=1$ and $\kappa_2=2+\epsilon$ as $\epsilon\to0$,
which does not become invalid, implies that our solution is the correct one.

To summarise, we have demonstrated that four Koopman decompositions are required to describe the interaction of a pair of solitons in
the KdV equation.
Each expansion is convergent in a particular region of space-time, either: (i) upstream of both solitons, (ii) downstream of both solitons,
(iii) between the solitons with the slower wave upstream of the faster or (iv) between the solitons with the faster wave upstream of the slower.
There is a simple logic to selecting the eigenfunctions required for any given expansion:
Alone, any individual soliton has a pair of Koopman decompostions;
an expansion describing the solution upstream of the soliton requires exponentially growing eigenfunctions
while temporally decaying eigenfunctions are needed downstream.
In the two-soliton interaction, this continues to apply.
However, products of the two sets of eigenfunctions must also be included to account for interaction between the solitons.

The approach outlined above naturally extends to arbitrary numbers of solitons, where construction of a Koopman decomposition at any point in space
requires products of all the growing eigenfunctions for any solitons downstream of that point and all of the decaying eigenfunctions from the upstream solitons.
For $N$ solitons, this would involve the solution of $N$ recurrence relations similar to (\ref{eq:2sol1}, \ref{eq:2sol2}) simultaneously.
The existence of multiple Koopman decompositions which partition the spatiotemporal domain to describe the full solution to
a nonlinear PDE has important consequences for DMD, which we now examine.

\section{Dynamic mode decomposition} 
\label{sec:dmd}
\begin{figure}
    \centering
    \includegraphics[width=\columnwidth]{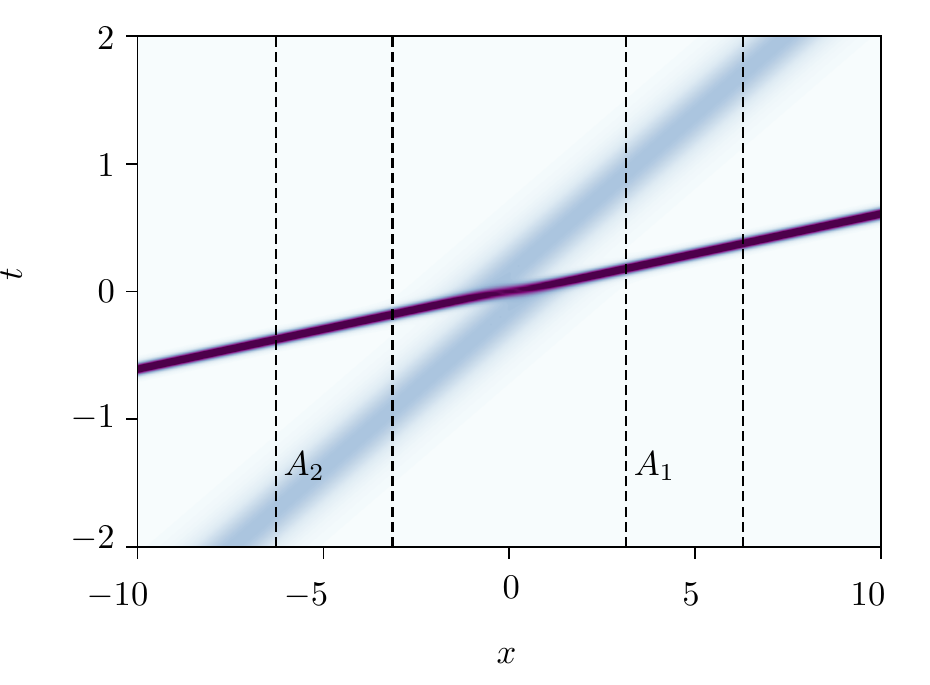}
    \caption{Two soliton solution to the KdV equation (\ref{eqn:2sol_ic}) visualized with contours of $-u$.
    Dashed lines identify DMD observation windows $A_1=(\pi,2\pi)$ and $A_2=(-2\pi,-\pi)$.}
    \label{fig:2sol_schem}
\end{figure}
\begin{figure}
    \includegraphics[width=\columnwidth]{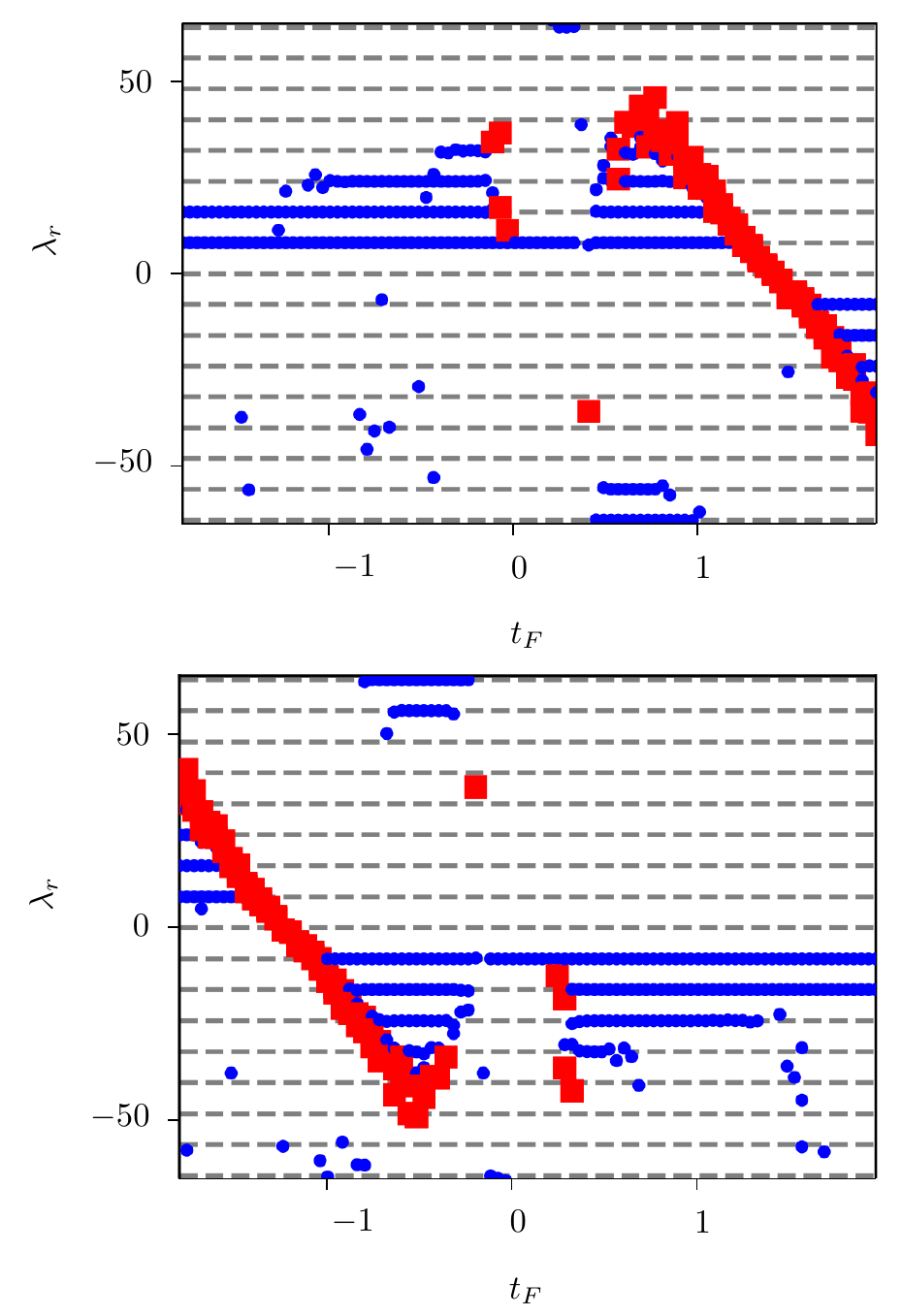}
    \caption{Real part of eigenvalues obtained in DMD calculations with a windowed observable $\boldsymbol g(u) = \mathbf u(x\in A_i)$ against the end time, $t_F$, of each DMD computation. Each DMD calculation is performed within a time window of length $T_w=0.4$ with snapshots available at a resolution of $\delta t=0.005$. The DMD timestep separating snapshots is $\delta t_{DMD}=0.01$ and $M=50$ snapshot pairs are used. (Top) observation window $A_1$, (bottom) observation window $A_2$.
    Note that blue circles identify purely real eigenvalues, red squares are complex conjugate pairs.}
    \label{fig:2sol_A1}
\end{figure}
Dynamic mode decomposition (DMD) can be an effective way to extract Koopman eigenvalues, modes and eigenfunctions from numerical data.
A rigorous connection between Koopman decompositions and DMD has been established under certain conditions \citep{Williams2015,Rowley2017}.
The key requirements are (i) that the Koopman eigenfunctions can be expressed as a linear combination of the elements of the DMD observable vector, $\{g_i(u)\}$, and (ii) that sufficient data is available.

A variety of methods have been proposed to augment DMD and aid its ability to extract Koopman eigenfunctions from data.
For example, in `extended' DMD, the observable vector $\boldsymbol g$ is built from a dictionary of functionals of the state.
For the nonlinear PDEs considered in this paper, we will see that standard DMD (where the observable is simply the state variable itself, $g_i = u(x=x_i)$), is sufficient to perform numerical Koopman decompositions, provided that the observations are restricted to a particular region of space-time where a single Koopman decomposition holds.

As a first example, consider the two-soliton KdV dynamics in figure \ref{fig:2sol_schem}.
The parameters match those considered in \S3.
Two groups of DMD calculations are considered with a windowed observable
\begin{equation}
    \boldsymbol g(u) = \mathbf u(x\in A_j),
\end{equation}
where the elements of $\mathbf u$ are observations of the state $u$ at the grid points, $(\mathbf u)_i := u(x=x_i)$,
and the choices for the window $A_j$ are identified in figure \ref{fig:2sol_schem}.
The DMD methodology is as described in \citep{Tu2014}.

For each of the two observation windows $A_j$, we perform many DMD calculations over short time intervals $T_w=0.2$.
The real parts of the eigenvalues obtained in these calculations are reported in figure \ref{fig:2sol_A1}, as a function of the final time of each individual DMD computation.
For the window $A_1$, while $t_F \lesssim 0$, the DMD identifies eigenvalues $\lambda_n = 8n, \; n\in \mathbb N$.
This corresponds to the analytical prediction for the Koopman decomposition upstream of both solitons,
where the set of Koopman eigenvalues required to correctly describe the time evolution is the product of the unstable eigenvalues associated with each individual soliton. 

Near $t_F = 0$, complex-conjugate pairs of eigenvalues (shown in red in figure \ref{fig:2sol_A1}) emerge and DMD is unable to find a robust representation that remains consistent between subsequent calculations.
This behaviour coincides with the observation window viewing regions of the solution which are expressed in terms of multiple Koopman decompositions; namely the top of the faster soliton is included in the observation window.
In this case, DMD is unable to build a consistent linear representation for the dynamics.

When $0.5 \lesssim t_F \lesssim 1$, the observation windows occupy a region of space-time between the two solitons, and the DMD algorithm is able to correctly identify the exponentially growing and decaying eigenvalues required in one of the central Koopman decompositions.
As well as the exponentially growing terms associated with being upstream of the slower soliton, $\lambda_n = 8n, \; n\in \mathbb N$, the rapidly decaying eigenvalue $\lambda_n=-64$ is also obtained.
This is the slowest-decaying eigenvalue associated with being downstream of the faster soliton.
Note that the other visible decaying eigenvalue ($\lambda_n=-56$) in this region is associated with the product of the first unstable Koopman eigenfunction associated with the slower soliton and the first stable Koopman eigenfunction connected with the faster wave, $\varphi_8(u)\varphi_{-64}(u)$ (see \S3).
Other decaying eigenvalues $\lambda_n = -8n \; n\in \mathbb N$ are also anticipated based on interactions $\varphi_8^j\varphi_{-64}^k$, though these terms are all much smaller in amplitude and are not picked up by the DMD.
These results are quickly contaminated with pairs of complex-conjugate modes that are associated with the appearance of the second crossover point -- the top of the slower soliton -- in the observation window.
Finally, towards the end of the later-time DMD calculations for window $A_1$, DMD starts to recover the purely decaying Koopman eigenvalues associated with the expansion downstream of both solitons.

Similar behaviour is observed for observation window $A_2$, which also shows evidence of three expansions.
In this instance, the eigenvalues identified between the solitons are similar to those seen for window $A_1$, but appear to be flipped about $\lambda_r=0$ as the observation window is upstream of the faster solution and downstream of the slower wave.
Therefore, while the upstream-of-both and downstream-of-both results are unchanged, the Koopman decomposition between the the two solitons involves the product of the unstable eigenvalues associated with the faster soliton and the stable eigenvalues of the slower pulse, i.e. the opposite of window $A_1$.

These observations suggest that the use of a spatially-restricted observable is a sensible choice in nonlinear problems involving spatially-localised dynamics.
This observable choice will allow individual Koopman eigenvalues and modes to be extracted by avoiding the inclusion of crossover points between multiple decompositions, for which DMD is unable to build a consistent linear operator.
In order to demonstrate the utility of such an approach, we examine a solution of the sine-Gordon equation,
\begin{equation}
    \partial_t^2 u =  \partial_x^2u - \text{sin}u,
\end{equation}
which arises in a variety of physical situations, including the propagation of dislocations through a crystal
and as a unitary theory for elementary particles \citep{scott1973soliton}.
Though analytical solution of the sine-Gordon equation is possible via the inverse scattering method \citep{ablowitz1973method}, we do not attempt to analytically find Koopman decompositions.
Instead, we will use the rules of thumb developed above for KdV to use DMD to identify Koopman eigenvalues.

\begin{figure}
    \centering
    \includegraphics[width=\columnwidth]{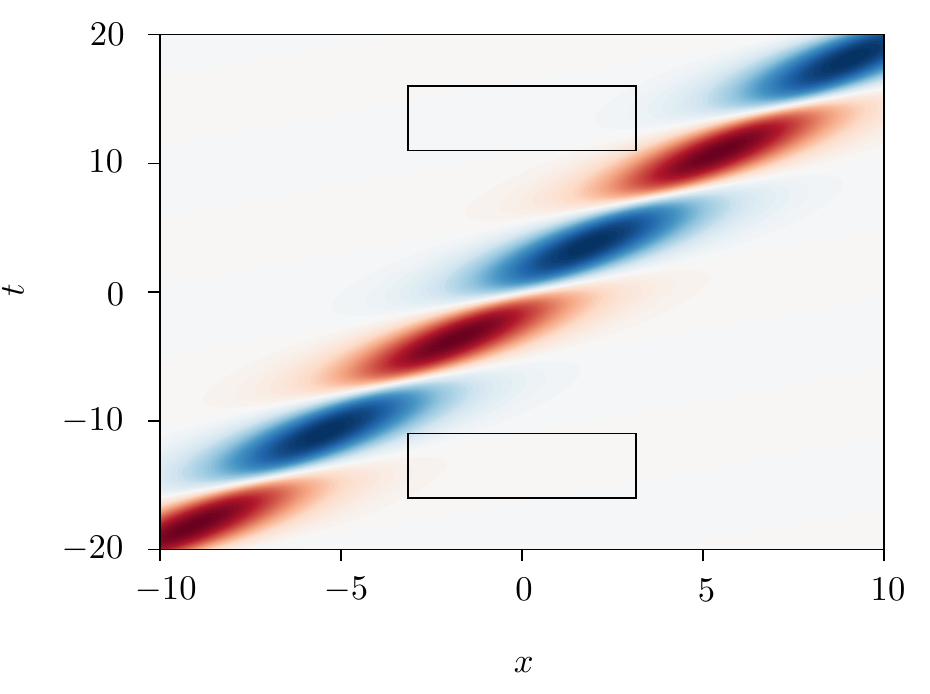}
    \caption{Moving breather solution to the sine-Gordon equation (\ref{eqn:breather}). Contours of $u$, with the observation windows for the DMD calculations in figure \ref{fig:sg_DMD} identified by black boxes.}
    \label{fig:sg_schem}
    \includegraphics[width=\columnwidth]{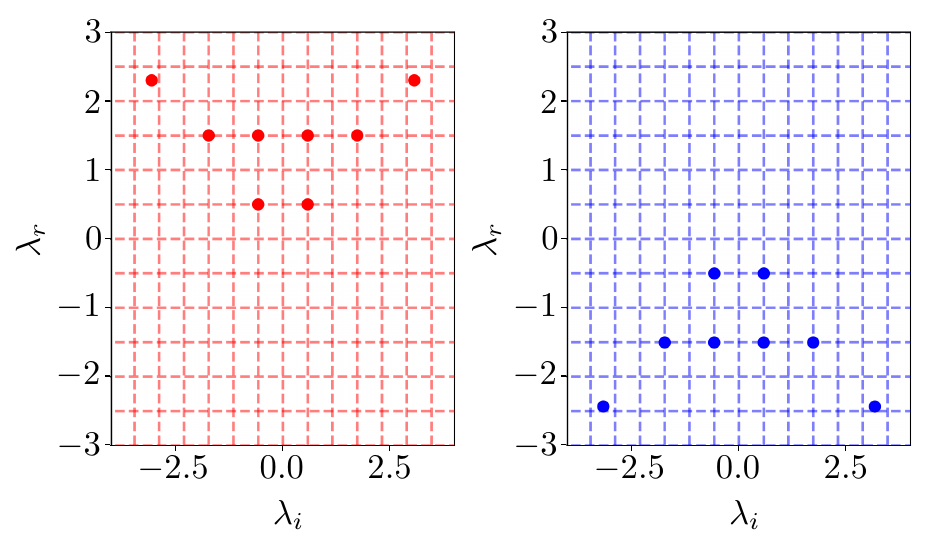}
    \caption{DMD applied to the sine-Gordon upstream (left) and downstream (right) of the breather (see figure \ref{fig:sg_schem}). In each calculation the observable is the state vector for $x\in(-\pi,\pi)$ and the time window length $T_w=5$. $M=400$ snapshot pairs are used with $\delta t= 0.1$.}
    \label{fig:sg_DMD}
\end{figure}

As an example, we focus on the moving breather solution \citep{DrazinSoliton},
\begin{equation}
    u_b(x,t) = 4\text{arctan}\left[\frac{\sqrt{1-l^2}}{l}\frac{\text{sin}(\gamma l(t-Vx))}{\text{cosh}(\gamma\sqrt{1-l^2}(x-Vt))}\right],
    \label{eqn:breather}
\end{equation}
where $\gamma:=1/\sqrt{1-l^2}$.
This solution is shown in figure \ref{fig:sg_schem} for $l=V = 1/2$, and is a localised relative periodic orbit.

Based on our analysis of both the Burgers and KdV equations, we anticipate the existence of a pair of Koopman decompositions upstream/downstream of the breather in terms of exponentially decaying/growing eigenvalues respectively.
In order to extract these representations, we conduct a pair of DMD computations with our observations restricted to windows upstream or downstream of the breather (marked in figure \ref{fig:sg_schem}).

The output of these calculations is reported in figure \ref{fig:sg_DMD}.
As anticipated, the calculations produce robust results both upstream and downstream of the oscillating pulse in terms of (temporal) exponential growth and decay.
Note that, unlike the Burgers and KdV equations, the eigenvalues are complex.
The upstream and downstream spectra are related via a reflection through $\lambda_r=0$.

Similarly to the one soliton solution of KdV, there is a natural interpretation of these results that is connected to the existence
of crossover points along connecting orbits between simple invariant solutions in state space.
In a co-moving coordinate, the moving breather may be interpreted as a homoclinic orbit about the trivial solution $u=0$,
and the DMD calculations identify the Koopman decompositions associated with the `repelling' and `attracting' halves of this trajectory.

%

\subsection{Periodic computational domains}

\begin{figure}
    \centering
    \includegraphics[width=\columnwidth]{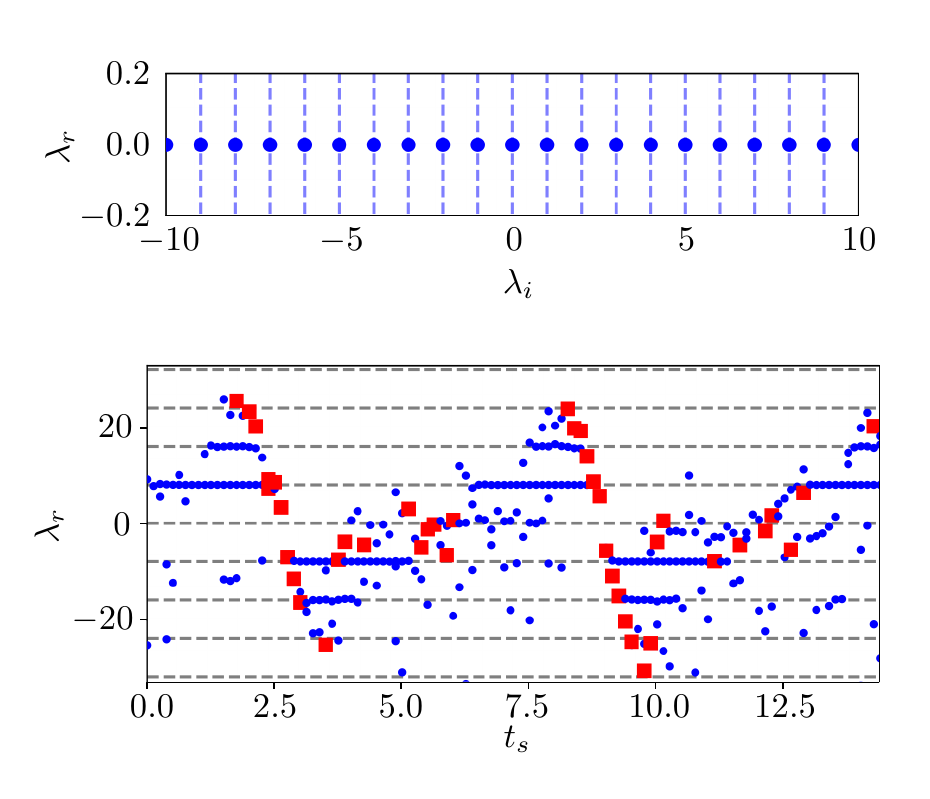}
    \caption{Two alternative DMD computations for the `one soliton' solution of the KdV equation evolving in a periodic computational domain of length $L=8\pi$.
    Top: full (unwindowed) state observable, $\boldsymbol g = \mathbf u$, observed over a time window $T_w=15$ with $M=400$ snapshot pairs. Vertical dashed lines identify multiples of the first non-zero frequency ($\omega=1$).
    Bottom: windowed state observable, $\boldsymbol g=\mathbf u(x\in A)$, where $A=(7\pi/2,4\pi)$. Multiple DMD computations are performed with time window length $T_w=0.2$ and the real part of the DMD eigenvalues are plotted against the start time of their respective DMD calculation. $M=40$ snapshot pairs are used. Throughout, $\delta t=0.0125$.}
    \label{fig:periodic_DMD}
\end{figure}
All of the problems studied so far in this work have been classical analytical solutions of integrable nonlinear PDEs on infinite domains.
However, studies of localised solutions to more complex systems (e.g. the Navier-Stokes equations \citep{Schneider2010}) are conducted in large
periodic computational domains.
As pointed out by \citet{Sharma2016}, Koopman decompositions for exact coherent structures in
spatially-periodic problems naturally take the form of travelling waves
and the (temporal) Koopman eigenvalues should all be purely imaginary.
This should be contrasted with the Koopman decompositions presented in this paper, which have all involved Koopman eigenvalues with non-zero real part.

To examine the connection between the assertions of \citet{Sharma2016} and the analytical Koopman decompositions
derived in this paper, we consider again the one-soliton solution to the KdV equation (see \S \ref{sec:kdvintro} and \S \ref{sec:1sol}).
Here, we supply the soliton $u = -2\text{sech}^2x$ as an initial condition in a numerical simulation where the KdV equation is solved numerically on a periodic domain of length $8\pi$.
A Fourier transform is applied in $x$; the nonlinear terms are evaluated in physical space before the transform is applied.
For time advancement, explicit Adams-Bashforth is used for the nonlinear terms and implicit Crank-Nicolson is used for the dispersive term.
The domain is long enough such that the error between the periodic numerical simulation and the true one-soliton solution,
$\| u_{per} - u_{sol} \|_2/ \|u_{sol}\|_2$, is about
$4 \times 10^{-4}$
after $\gtrsim 3$ flow-through times.

In figure \ref{fig:periodic_DMD} we report the results of two sets of DMD calculations on this one soliton KdV evolution.
In the first, a single computation, we perform standard DMD on the full state vector (i.e. over the entire spatial domain) for a time window spanning many flow-through times.
As anticipated, the DMD eigenvalues are all purely imaginary and are multiples of a fundamental harmonic $\omega=1$ (on this domain the flow-through
time of the isolated soliton is $T=2\pi$).
The DMD modes (not shown) are Fourier modes.

In the second set of calculations, we adopt the approach we have advocated for the infinite domains. We perform DMD
on a windowed observable $\boldsymbol g(u) = \mathbf u(x\in A)$, where $A = (7\pi/2, 4\pi)$,
conducting a sequence of DMD calculations on short time windows $T_w=0.2$.
The real part of the eigenvalues obtained in each calculation are shown in the lower panel
of figure \ref{fig:periodic_DMD}.
As the soliton repeatedly passes through the domain, the DMD calculations continually pick up the upstream/downstream eigenvalues associated with the solution on an unbounded domain (i.e. one of $\lambda_n = \pm 8n$, $n\in \mathbb N$).

In this problem the ``correct'' decomposition is the one involving purely imaginary eigenvalues, regardless of domain length (as long as it remains finite).
This can be demonstrated explicitly by considering the periodic `cnoidal' solution of the KdV equation \citep{korteweg_de_vries},
\begin{equation}
    \label{eq:cnoidal}
    u(x,t)=-2m\, \mathrm{cn}^2\left(x-(8m-4)t\right),
\end{equation}
where $\mathrm{cn}$ is the Jacobi elliptic cosine function with modulus $m\in [0,1]$.
Equation (\ref{eq:cnoidal}) is a right-moving travelling wave (phase speed $8m-4$),
and is spatially periodic
with period $2K$, where $K=K(m)$ is the complete elliptic integral of the first kind \citep{abramowitz_stegun}.
As $m\to 0$, (\ref{eq:cnoidal}) becomes the small-amplitude solution to the linearised KdV, a pure cosine.
As $m\to 1$ however, the peaks become repeated copies of the one-soliton solution, very widely separated in $x$:
on any finite spatial interval at fixed $t$, (\ref{eq:cnoidal})$\to$(\ref{eq:1sol}) as $m\to1$.

The Fourier series for (\ref{eq:cnoidal}) can be calculated using the series for $\mathrm{dn}^2$
given by \citet{oberhettinger1973fourier} and the identity $\mathrm{dn}^2(x)=1-m+m\,\mathrm{cn}^2(x)$,
giving
\begin{multline}
    \label{eq:cnoidalkmd}
    u(x,t)=-2\left(\frac{E}{K}+m-1\right)\\-\frac{4\pi^2}{K^2}\sum_{n=1}^\infty \frac{nq^n}{1-q^{2n}} \cos\left(\frac{n\pi}{K}\left\{x-(8m-4)t\right\}\right),
\end{multline}
where $E$ is the complete elliptic integral of the second kind, and $q(m)=e^{-\pi K(1-m)/K(m)}$ is the `nome' \citep{abramowitz_stegun}.

Viewing (\ref{eq:cnoidalkmd}) as a Koopman mode decomposition by writing the cosine in terms of exponentials,
we identify Koopman eigenvalues $i n\pi(8m-4)/K$ for $n\in\mathbb{Z}$.
These are purely imaginary
(or zero), as anticipated from periodicity, and should be contrasted to the purely real
Koopman eigenvalues found for the single soliton in isolation (\ref{eq:1sol}).

Despite the correspondence between the one-soliton solution to KdV and the limiting form of the periodic cnoidal wave,
the isolated soliton Koopman decomposition is not obtained in the large-domain limit due to the fact that
an \emph{infinite} domain is required to obtain the scattering data that define the Koopman eigenfunctions.
Furthermore, in contrast to Koopman decompositions constructed in \S \ref{sec:kdv} for solitons on infinite domains,
the Koopman modes and eigenvalues obtained in this periodic example are dependent on the domain length
rather than being purely tied to the soliton itself.
There are additional numerical issues too -- the periodic Koopman decomposition can be difficult to obtain in
the large-domain limit since very many Fourier modes
are required to resolve the evolution (see Appendix).

The striking difference between the periodic Koopman decomposition and the decomposition for a truly localised structure
is somewhat disconcerting, since simulations of localised structures are often conducted on large periodic domains under
the assumption that the true isolated structure is well approximated.
However, the windowed DMD results reported in figure \ref{fig:periodic_DMD} indicate that the Koopman decompositions for the
localised structure can still be obtained in periodic computations by using a spatially localised observable.
These results suggest that the two alternative strategies for DMD are both equally valid, depending on what the computation is
designed to find:
(i) the `standard' approach using the full state vector which will identify purely imaginary, domain-dependent Koopman eigenvalues (if
the structure is allowed to pass through the entire domain) and
(ii) the windowed observable which can identify the growing/decaying Koopman eigenvalues associated with upstream/downstream expansions
for a truly localised structure.


\section{Conclusions}
\label{sec:conclusion}
%
%

In this paper, we have derived Koopman decompositions in a number of problems involving the propagation and interaction
of isolated structures, namely a front in the Burgers equation and solitons in the KdV equation.
The results indicate that isolated nonlinear waves require two Koopman decompositions to describe their evolution, which converge
either upstream of downstream of the structure.
In many-soliton interactions, multiple Koopman decompositions are required, and selecting the convergent expansion
at any point requires knowledge of the relative positions of all solitons (i.e. whether they are upstream or downstream of the observation point).

We proposed a simple modification to the standard DMD methodology that allows allows the algorithm to identify the individual
Koopman decompositions around the isolated structures.
This approach was used to identify the various Koopman decompositions in a two-soliton interaction solution of KdV,
before we applied it to the sine-Gordon equation where the analytical eigenvalues are at present unknown.
The results suggest that the need for multiple Koopman decompositions to cover the full spatio-temporal domain
may be a generic feature of nonlinear PDEs.

Further work is required to assess the extent to which these results extend to more complex systems, such as the full Navier-Stokes equations.
As a starting point, the windowing approach could be applied to some of the known localised relative periodic solutions in pipe flow \citep{Avila2013}.
In addition, our analysis of the KdV equation was restricted to pure soliton evolution -- i.e. dispersive effects were absent.
The inclusion of dispersion will introduce a continuous spectrum of purely imaginary Koopman eigenvalues.
It would be of interest to know how the presence of these effects impacts the capability of DMD to identify the eigenvalues associated with
the coherent structures, and whether some of the recent proposed modifications to the algorithm, such as augmenting the observable with
other functionals, can help.


\appendix

\section{Further details on the cnoidal wave}
In this appendix we briefly discuss the behaviour of the Koopman decomposition for the cnoidal wave (equation \ref{eq:cnoidalkmd})
in the large-domain limit.

In the limit $m \to 1$, the elliptic integral $K(1-m)\to \pi/2$, so $q \sim e^{-\pi^2 /2K}$ and
\begin{equation}
    K\sim -\frac{\pi^2}{2\log q}.
\end{equation}
Therefore the $n$th Fourier coefficient from (\ref{eq:cnoidalkmd}) obeys
\begin{equation}
    -\frac{4\pi^2}{K^2} \frac{nq^n}{1-q^{2n}} \sim -\frac{16 \left(\log q\right)^2}{\pi^2} \frac{nq^n}{1-q^{2n}}.
\end{equation}
Since $q\to1$ as $m\to1$, we expand with $\epsilon=1-q$ to give
\begin{align}
    -\frac{4\pi^2}{K^2} \frac{nq^n}{1-q^{2n}}
    \sim -\frac{16 \left(-\epsilon\right)^2}{\pi^2} \frac{n}{2n\epsilon} \to 0.
\end{align}
Since every Fourier coefficient approaches $0$ as $m\to 1$, but the cnoidal wave peaks tend to a fixed height of $-2$,
an increasing number of Fourier modes (which are Koopman modes here) must be used to approximate the solution.
This means that for very isolated solitons in a periodic domain, a large number of DMD modes will be required.

\begin{acknowledgments}
    The authors wish to thank C. P. Caulfield and R. R. Kerswell for much helpful advice.
    This work is supported by EPSRC Programme Grant EP/K034529/1
    entitled `Mathematical Underpinnings of Stratified Turbulence’.
\end{acknowledgments}

\bibliography{pp19}

\end{document}